\documentclass[twocolumn,showpacs,preprintnumbers,amsmath,amssymb]{revtex4}

\usepackage{graphicx}
\usepackage{amssymb}
\usepackage{dcolumn}
\usepackage{bm}

\begin{document}


\title{Origin of Universal Optical Conductivity and Optical Stacking Sequence Identification in Multilayer Graphene}
\author{Hongki Min$^{1,2,3}$}
\author{A. H. MacDonald$^{3}$}
\affiliation{
$^{1}$Center for Nanoscale Science and Technology, National Institute of Standards and Technology, Gaithersburg, MD 20899-6202\\
$^{2}$Maryland NanoCenter, University of Maryland, College Park, MD 20742\\
$^{3}$Department of Physics, University of Texas at Austin, Austin TX 78712
}

\date{\today}

\begin{abstract}
We show that the origin of the universal optical conductivity in a normal $N$-layer graphene multilayer is an emergent chiral symmetry which guarantees that $\sigma(\omega)=N\sigma_{uni}$ in both low and high frequency limits. [$\sigma_{uni}=(\pi/2) \, e^2/h$].  We use this physics to relate intermediate frequency conductivity trends to qualitative characteristics of the multilayer stacking sequence.
\end{abstract}

\maketitle

\noindent
{\em Introduction.}---Graphene is an atomically two-dimensional material which can be viewed 
either as a single sheet of graphite or as a large unrolled nanotube. 
Experimenters\cite{exp1,exp2,exp3,exp4,review} have recently made progress in preparing and 
measuring the electronic properties of single and multilayer 
graphene sheets.  One particularly intriguing property of neutral single-layer 
graphene sheets is its interband optical conductivity which is
expected\cite{Ando,Gu,Fa1} to be approximately constant over a broad range of frequencies with a value
close to 
\begin{equation}
\sigma_{uni} = \frac{\pi}{2}\frac{e^2}{h},
\end{equation} 
dependent only on fundamental constants of nature.
Experiments\cite{GeimScience,CrommieScience,Basov,Heinz} have demonstrated that 
corrections to the constant universal conductivity, which might be 
expected to follow from electron-electron interactions\cite{He,Mi,Sachdev} or
refinements of the Dirac-equation band-structure model\cite{stauber} for example, are small. 
Recently Gaskell {\em et al.}\cite{gaskell2008} found that for frequencies in the optical range the  
conductivity per layer in multilayer graphene sheets is also surprisingly close to  $\sigma_{uni}$.
Separately Kuzmenko {\em et al.}\cite{Kuzmenko_graphite} demonstrated experimentally that in bulk graphite the
optical conductivity per layer has a smoother frequency dependence and is even more uniformly
close to $\sigma_{uni}$ than in thin multilayers, and explained the weak
frequency dependence they found in terms of the crossover to three-dimensions.  
In this Letter we identify the emergent chiral symmetry of multilayers\cite{min2008a,min2008b} as 
a key element of the physics responsible for the 
ubiquity of $\sigma_{uni}$ in multilayer graphene systems.  
 
A single graphene sheet consists of a honeycomb lattice of carbon atoms.  
Graphene $\pi$-orbitals can be viewed\cite{review} as possessing a ``which sublattice"
pseudospin degree of freedom.  The envelope functions of their electron waves are described by a two-dimensional 
massless Dirac equation which possesses pseudospin chiral symmetry and leads to eigenspinors
in which the phase difference between sublattices $\phi= \pm J \phi_{{\bm k}}$ where the pseudospin chirality $J=1$ and $\phi_{\bm k}=\tan^{-1}(k_y/k_x)$. 
The pseudospin chirality is defined by this equation as rate at which pseudospin orientation varies with momentum orientation.

When sheets are stacked to form a multilayer system there is an energetic preference for an arrangement 
in which each layer is rotated by 60$^\circ$ with respect to one of the two
sublattices of its neighbors. This prescription generates three distinct planar projections of the 
honeycomb lattice (A, B, and C) and therefore $2^{N-2}$ distinct $N$-layer sequences.
We refer to multilayers in this class as {\em normal}.
Repeated AB (Bernal) stacking and repeated ABC (orthorhombic) stacking should be viewed as extreme cases,
as we explain in more detail below. 
The emergent chiral symmetry we discuss below applies for any normal multilayer.
AA (hexagonal) stacking (placing a layer directly on top of another) is energetically costly\cite{simplehexagonal},
does not yield chiral symmetry, and has clear optical signatures.
Our discussion of optical properties for normal multilayer graphene systems starts from a model, referred to below as the {\em ideal} model, 
in which only the dominant interlayer nearest-neighbor hopping processes couple individual-sheet Dirac-equation waves.

The optical conductivity of an $N$-layer system is expected to approach $N\sigma_{uni}$ for frequencies that exceed the interlayer-coupling scale
but are smaller than the $\pi$-bandwidth scale,  
since the layers then contribute independently and the Dirac model still applies.
The ideal model of normal graphene multilayers has a surprising property which we have explained previously\cite{min2008a,min2008b}.
In the low-energy limit its spectrum separates asymptotically into $N_{D} \le N$ decoupled pseudospin doublets,
each of which has chiral symmetry with a chirality $J$ which can in general be larger than $1$.
We demonstrate below that the conductivity of a pseudospin doublet with chirality $J$ is $J \sigma_{uni}$.
It then follows from the chirality sum rule\cite{min2008a}, 
\begin{equation}
\label{eq:sumrule}
\sum_{n=1}^{N_{D}} \; J_n = N, 
\end{equation} 
that the conductivity of the ideal model unexpectedly also approaches $N \sigma_{uni}$ in the $\omega \to 0$ limit.  
Since the asymptotic pseudospins which emerge at low-energies are in general spread 
across a number of different layers and are qualitatively dependent on the way in which the layers are stacked, 
the low-frequency limit of the interband conductivity does not result from independent single-layer
contributions but has a completely different origin.  We show below that corrections to the ideal model
are small and therefore use it to analyze correlations between stacking and deviations from $N \sigma_{uni}$ at 
intermediate frequencies.
 
\noindent
{\em Chiral doublet conductivity.}---The Kubo formula for the real part of the optical conductivity, $\sigma_R(\omega)\equiv {\rm Re}[\sigma_{xx}(\omega)]$, of an 
$M$-band two-dimensional electron-gas system is
\begin{eqnarray}
\label{eq:conductivity}
\sigma_R(\omega)=&-&{\pi e^2 \over h}\sum_{n\neq n'}\int {d^2 k \over 2\pi} \; {f_{n,\bm{k}}-f_{n',\bm{k}} \over \epsilon_{n,\bm{k}}-\epsilon_{n',\bm{k}}}  \\
&\times&\left|\left<n,\bm{k}\right|\hbar v_x\left|n',\bm{k}\right>\right|^2  \; \delta( \hbar\omega+\epsilon_{n,\bm{k}}-\epsilon_{n',\bm{k}}), \nonumber
\end{eqnarray}
where $\epsilon_{n,\bm{k}}$ and $\left|n,\bm{k}\right>$ are eigenvalues and eigenvectors of the $M \times M$ Hamiltonian matrix ${\cal H}$, 
$f_{n,\bm{k}}$ is a Fermi occupation factor and $v_a = \partial {\cal H}/\hbar \partial k_a$ is the velocity operator.  
The $M=2$ Hamiltonian matrix of a doublet with chirality $J$ is 
\begin{equation}
{\cal H}_J=\gamma_1 \; \left(
\begin{array}{cc}
0 &(\nu_{\bm{k}}^{\dagger})^J \\
(\nu_{\bm{k}})^J &0 \\
\end{array}
\right),
\end{equation}
where $\gamma_1$ is the nearest-neighbor interlayer hopping, $\nu_{\bm{k}}\equiv\hbar v k e^{i\phi_{\bm{k}}} /\gamma_1$ and $v$ is the effective in-plane Fermi velocity (for example, $v={\sqrt{3}\over 2}{a\gamma_0\over \hbar}$ for $J=1$ monolayer and $J=2$ bilayer graphene where $\gamma_0$ is the nearest-neighbor intralayer hopping and $a=2.46$ $\rm \AA$ is the graphene lattice constant).

This chiral-invariant Hamiltonian has negative ($s=-1$) and positive ($s=1$) energy eigenstates 
with eigenenergies  
$\varepsilon_{s,\bm{k}}=s \gamma_1 |\nu_{\bm{k}}|^J$ and eigenvectors 
\begin{equation}
\left|s,\bm{k}\right>=
{1 \over \sqrt{2}}\left(
\begin{array}{c}
s \\
e^{i J\phi_{\bm{k}}}\\
\end{array}
\right).
\end{equation}
It follows that the interband matrix element of the velocity operator is
\begin{equation}
\left<s,\bm{k}\right|\hat{v}_x\left|-s,\bm{k}\right> \; =i \; J v s |\nu_{\bm{k}}|^{J-1} \sin \phi_{\bm{k}}.
\end{equation}
Inserting these expressions into the Kubo formula, multiplying the result by a factor of 
$g_s g_v = 4$ to account for the spin and valley degeneracy of graphene systems,
we obtain 
\begin{equation}
\label{eq:sigmaJ}
\sigma_R(\omega) \; = \; {J\pi \over 2}  {e^2\over h} \; = \; J \sigma_{uni}.
\end{equation} 
Since $\gamma_1$ is the only energy scale in the Hamiltonian, it is clear prior to calculation, that $\sigma_R(\omega) \propto (e^2/h) (\hbar\omega/\gamma_1)^{\ell}$.
The velocity matrix element, joint density-of-states, and energy denominator factors combine 
so that $\ell=0$ for every value of $J$ and, importantly, so that $\sigma_R(\omega) \propto J$.    

\noindent
{\em Ideal-model conductivity.}---It follows from the chiral doublet conductivity Eq.~(\ref{eq:sigmaJ}),
the emergent chiral symmetry\cite{min2008a,min2008b} of the ideal model, and the chirality sum rule Eq.~(\ref{eq:sumrule}),
that the ideal-model conductivity for normal $N$-layer graphene satisfies $\lim_{\omega \to 0} \sigma_R(\omega) = N \sigma_{uni}$. 
The numerical calculations necessary to evaluate $\sigma_R(\omega)$ at intermediate frequencies are also remarkably simple.
Because the band energies of the 
ideal model are dependent only on the magnitude of wavevector $\bm{k}$, the only $\phi$-dependent 
quantities which appear in the wavevector integral for the conductivity (Eq.~(\ref{eq:conductivity})) are the velocity matrix elements
$M(\phi)=\left<n,\bm{k}\right|\hbar v_x\left|n',\bm{k}\right>$. 
In an $N$-layer stack the fastest possible angle variation in any matrix element varies as $\exp(\pm i N \phi)$. It follows that all angle integrals are evaluated {\em exactly} by summing over $2N+1$ equally spaced orientations.
The wavevector magnitude integrals are performed by solving for $|\bm{k}|$ values at which 
interband energy differences are equal to $\hbar\omega$.  The ideal-model conductivity of any multilayer stack can be 
evaluated very accurately with a relatively small numerical effort. 

\begin{figure}[htb]
\begin{center}
\includegraphics[width=0.9\linewidth]{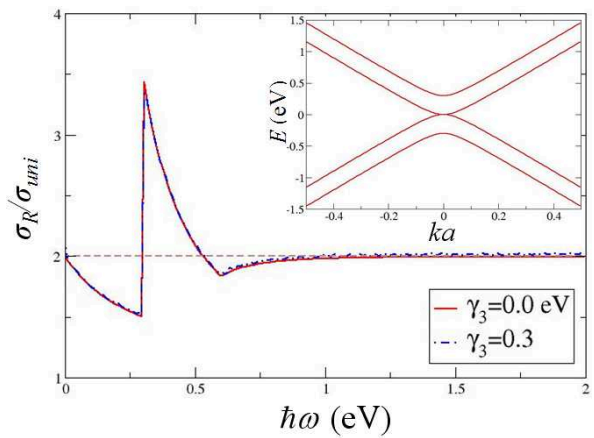}
\caption{(color online) Conductivity for ideal-model bilayer graphene
($\gamma_3=0$) and for a more realistic model with distant neighbor interlayer hopping ($\gamma_3=0.3$ eV). The inset shows ideal-model band structure of bilayer graphene.}
\label{fig:layer2}
\end{center}
\end{figure}

\begin{figure}[htb]
\begin{center}
\includegraphics[width=1\linewidth]{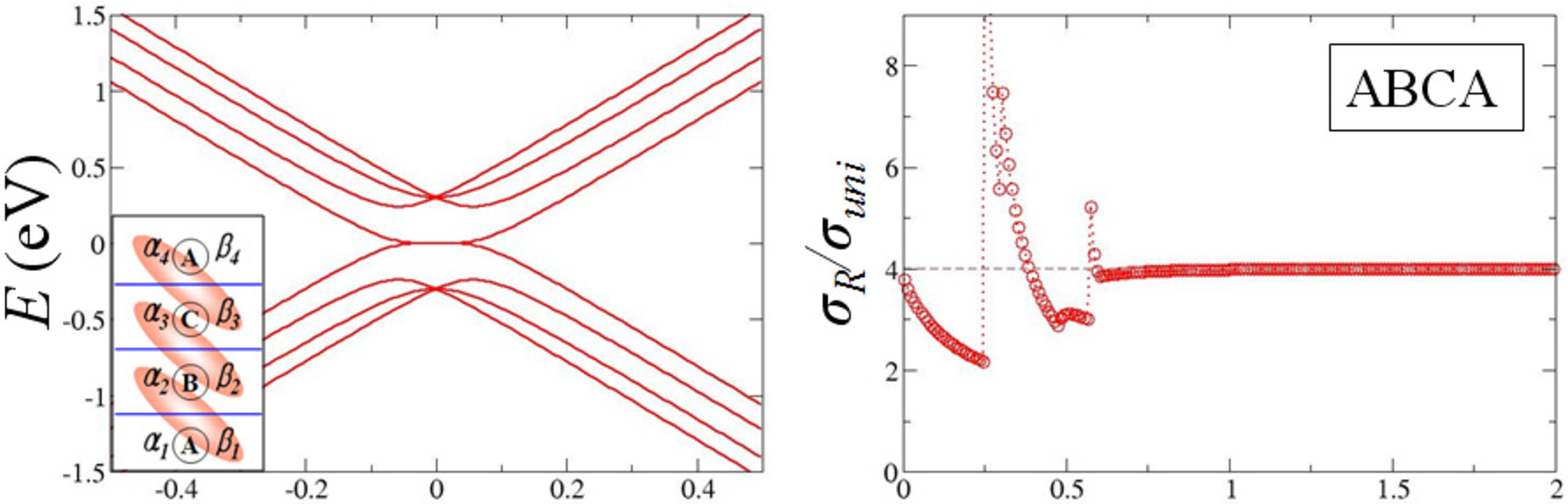}
\includegraphics[width=1\linewidth]{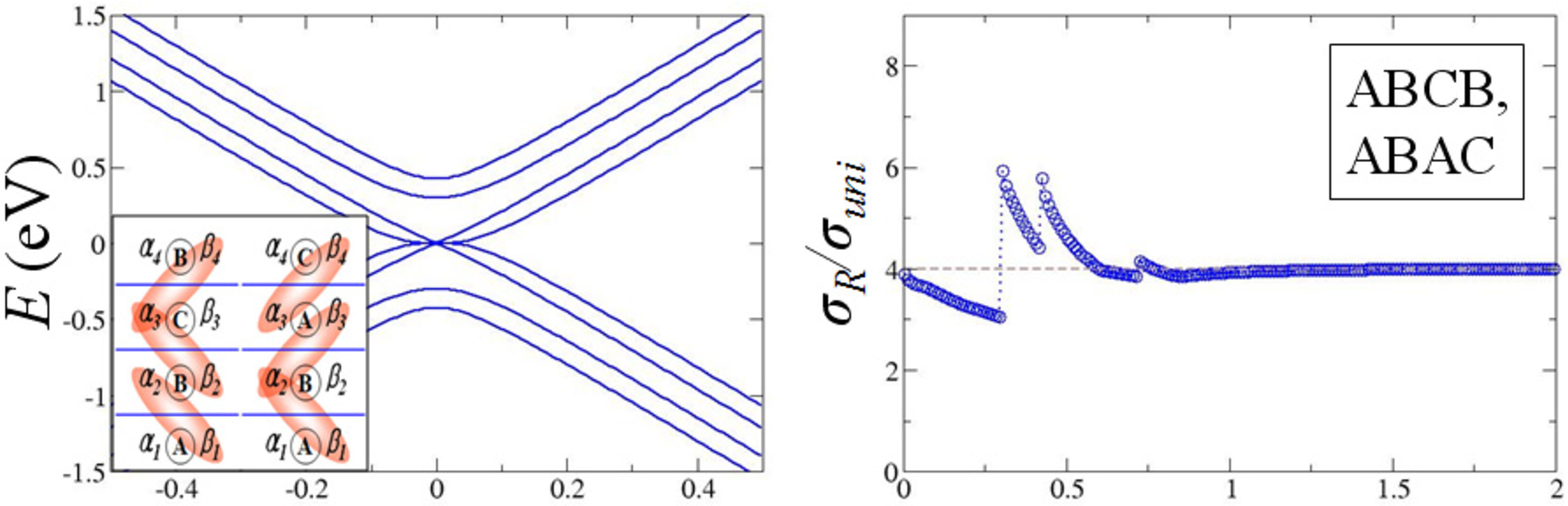}
\includegraphics[width=1\linewidth]{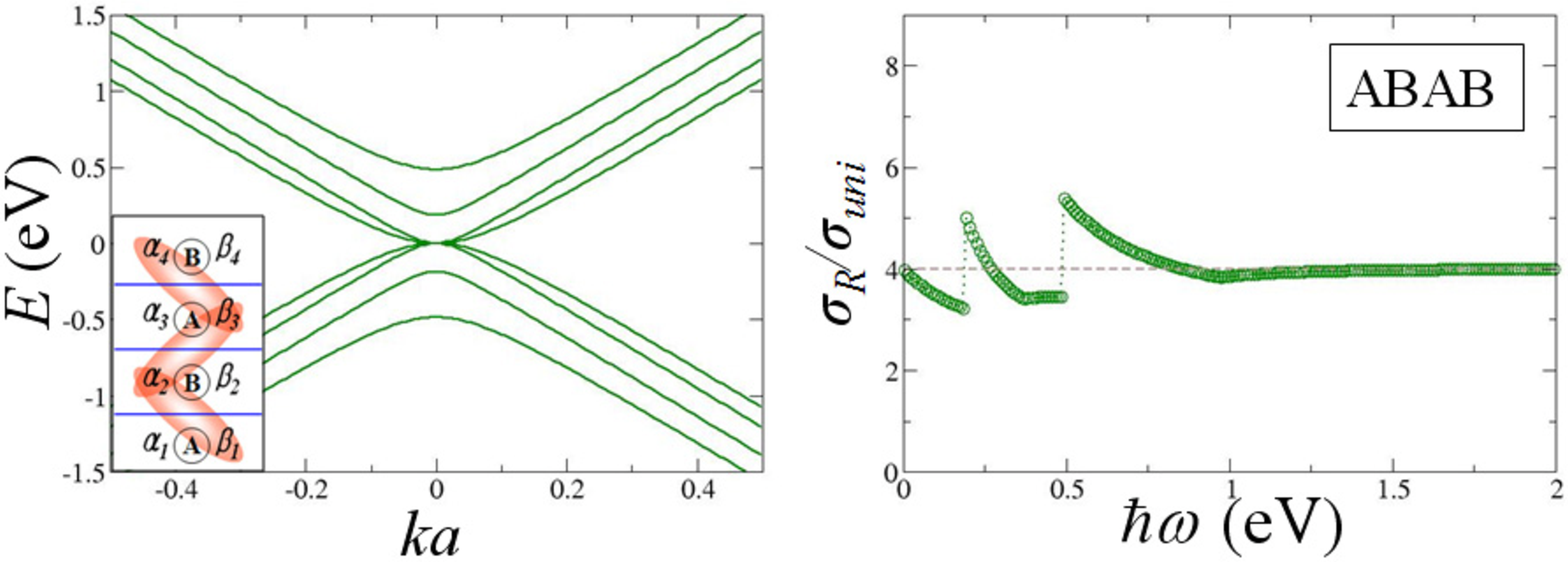}
\caption{(color online) Ideal-model band structure and real part of the conductivity for all tetralayer graphene stacks, ABCA (top), ABCB, ABAC (middle) and ABAB (bottom). The insets show stacking diagrams where shaded ovals link sublattices $\alpha$ and $\beta$ to the nearest interlayer neighbors.}
\label{fig:layer4}
\end{center}
\end{figure}

\begin{figure}[htb]
\begin{center}
\includegraphics[width=0.9\linewidth]{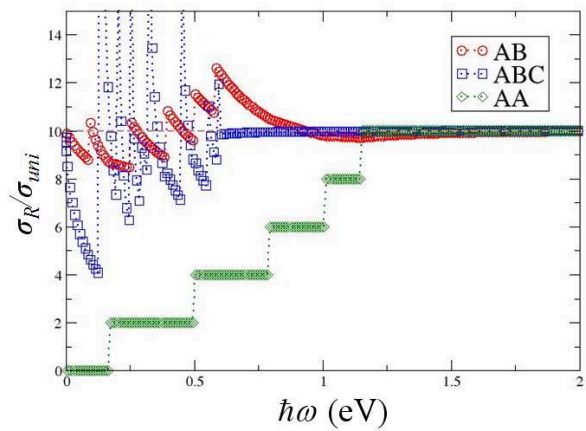}
\caption{(color online) Real part of the ideal-model conductivity for Bernal (AB), orthorhombic (ABC), and hexagonal (AA) 10 layer graphene stacks.}
\label{fig:layer10}
\end{center}
\end{figure}

We have evaluated $\sigma_R(\omega)$ curves for many stacking arrangements; all results 
confirm our claim that for the ideal model $\sigma_R(\omega) \to N \sigma_{uni}$ in low and high frequency limits.
In Figs.~\ref{fig:layer2}--\ref{fig:layer10}
we show representative results for some 
2, 4, and 10 layer stacks which allow us to discuss corrections to the ideal model 
and trends in the relationship between stacking and intermediate frequency deviations from $N\sigma_{uni}$.

\noindent 
{\em Discussion.}---
The optical conductivity of bilayers (Fig.~\ref{fig:layer2}) has 
been studied both theoretically\cite{Abergel,Nicol} and experimentally\cite{CrommieScience,basov_bilayer,kuzmenko_bilayer,Li_bilayer} in previous work. 
The electronic structure consists\cite{min2008a} of a $J=2$ chiral doublet at the Fermi energy
(the simplest example of emergent chiral symmetry) and a two-site chain split-off band. The low-frequency conductivity 
originates from transitions within the $J=2$ doublet.
Because the corresponding transition matrix elements are finite at $\bm{k}=0$,
the strongest infrared (IR) feature occurs at the onset energy, $\hbar\omega=  \gamma_1$, of transitions between the chiral doublet and the 
split-off bands.  
($\gamma_1 = 0.3$ eV\cite{review} in all our calculations.)
The feature associated with transitions from split-off valence to split-off conduction bands at $2 \gamma_1$ is weaker 
because in this case the velocity matrix elements vanish at the $\bm{k}=0$ onset.   
Inclusion of remote interlayer hopping $\gamma_3$, or of any other corrections to the 
ideal model has little influence on the optical conductivity.


The same patterns continue in thicker multilayers. The low-frequency conductivity $N \sigma_{uni}$ comes entirely from transitions within asymptotically free chiral doublets,
each of which is dispersed across the multilayer.  The high-frequency conductivity, also $N \sigma_{uni}$, comes from transitions within decoupled single-layers.  The crossover 
at intermediate frequencies is punctuated by the onset of a series of interband transitions, with the strongest features coming from 
transitions between the low-energy chiral doublets and split-off bands.  The $N=4$ case, illustrated in    
Fig.~\ref{fig:layer4}, has four distinct stacking sequences, two of which are related by inversion symmetry.  
The three inequivalent cases are orthorhombic ABCA stacking which yields\cite{min2008a} a $J=4$ low-energy chiral doublet
and three two-site-chain split-off bands,
Bernal ABAB stacking which yields\cite{min2008a} two $J=2$ chiral doublets and four-site-chain split-off bands, and intermediate ABCB stacking which yields\cite{min2008a} 
$J=3$ and $J=1$ chiral doublets and both three and two-site-chain split-off bands.
The optical conductivity of the orthorhombic ABCA stack has a divergent  
IR feature associated with $J=4$ chiral doublet to two-site chain transitions.  The onset of this absorption
band is below $\gamma_1$ because one of the three two-site-chain 
split-off bands disperses toward the Dirac point and 
has an extremum at finite $ka\approx 0.1$,
implying a divergent joint density of states.  The Bernal ABAB stack has two jump-discontinuity IR features
associated with ${\bm k}=0$ transitions between the $J=2$ doublets and $E_r = \pm 2 \gamma_1 \cos(r\pi/5)$ four-site-chain\cite{min2008a} split-off bands with $r=1,2$.
Similarly the ABCB stack has strong IR features associated with transitions between the chiral doublets and 
both $E=\pm\gamma_1$ two-site chain and $E = \pm\sqrt{2} \gamma_1$ three-site chain bands.

Normal graphene stacks can always be organized into Bernal and orthorhombic segments.  In general stacks with more 
orthorhombic segments have fewer\cite{min2008a} low-energy chiral doublet bands (which must therefore have higher chirality because 
of the chirality sum rule) and shorter\cite{min2008a} chain split-off bands.  Level repulsion among the short-chain bands 
tends to cause some to disperse toward the Dirac point and have finite $k$ extrema.  Stacks with more Bernal segments will have 
more chiral doublets with lower chirality and longer-chain split-off bands.  In all cases the strongest 
IR features are associated with transitions between chiral doublets and split-off bands.

Since long-chain states are spread
over a broader range of energies, stacks with more Bernal character tend to have IR features
that are weaker and spread over a wider energy range.  This trend is clear in  
Fig.~\ref{fig:layer10}, which compares the optical conductivities of 10-layer Bernal and orthorhombic stacks.
In the Bernal case\cite{Koshino} transitions from chiral doublets to 10-site-chain split off bands lead to jump
discontinuities in $\sigma_R(\omega)$ at $\hbar\omega = 2 \gamma_1 \cos(r \pi/11)$ for $r=1,2,3,4,5$.  In the orthorhombic case, the 
first IR feature appears at a larger frequency but the deviations from $10\sigma_{uni}$ are larger.    

Figure ~\ref{fig:layer10} also shows conductivity results for a 10 layer hexagonal stack, in which honeycomb layers are placed one 
on top of the other.  Stacks of this type do not satisfy the chirality sum rule\cite{min2008b} and invariably lead to $\sigma_R(\omega)$ curves which are 
suppressed at low frequencies.  
As illustrated in Fig.~\ref{fig:layer10}, even $N$ hexagonal stacks 
do not have a low-energy chiral doublet, thus $\sigma_R(\omega)$ vanishes at small $\omega$
and then increases toward $N \sigma_{uni}$ in steps of $2 \sigma_{uni}$.
For odd $N$ the stack has a single $J=1$ chiral doublet at low-energies so that $\sigma_R(\omega)$ starts at $\sigma_{uni}$ in the 
low-frequency limit and then increases in $2 \sigma_{uni}$ steps toward $N \sigma_{uni}$.  More generally,  
stacks with AA segments always have $\sigma_R(\omega)$ in the low-frequency limit smaller than $N \sigma_{uni}$
and approach this limit only at high-frequencies. 

The full electronic structure of graphene multilayers is usually discussed in terms of an appropriate adaptation of the 
Slonczewski-Weiss-McClure (SWM)\cite{SWM,review} parameterization of graphite's bands.  A SWM-type model improves the ideal model by 
accounting for small differences between the energies of $\pi$-orbitals on inequivalent carbon atoms and for 
various distant neighbor hopping amplitudes.  In the bilayer case the only important additional parameter is the 
distant neighbor interlayer hopping amplitude $\gamma_3$, and as noted earlier this process has negligible 
influence on $\sigma_R(\omega)$.  These refinements of the electronic structure model lead in general to
two-dimensional (2D) electron and hole Fermi surfaces with Fermi energies that are much smaller than the dominant interlayer coupling energy $\gamma_1$.  
The appearance of 2D Fermi surfaces therefore has little influence on the IR conductivity.  The presence of 2D Fermi surfaces (and in the 
limit of graphite of 3D Fermi surfaces) does imply that $\sigma_R(\omega)$ will in general have a small amplitude Drude peak, 
not accounted for in the present discussion.  The Drude peak will take a small amount of spectral weight\cite{fsum} from the IR 
interband transitions.  More realistic models also do in general break particle-hole symmetry.  This refinement 
will cause an IR feature associated with a particular split-off valence band to chiral doublet transition to appear at a slightly different
frequency than the corresponding transition in the ideal model.  These caveats notwithstanding, the origin of the generic $\sigma_R(\omega) \approx N \sigma_{uni}$ 
behavior in graphene multilayers is explained most succinctly by the ideal model, and in particular by its emergent chiral symmetry.

The ideal model is also able to capture the stacking structure implications of measured IR conductivity features:
the high-frequency decoupled layer $N \sigma_{uni}$ limit is approached for $\hbar\omega \gtrsim 2 \gamma_1 \approx 0.6$ eV for 
normal stacks, but only at higher frequencies $\approx 4 \gamma_1 \approx 1.2$ eV when AA stacking faults are present.
AA stacking is also indicated by suppressed conductivity at lower frequencies,
as discussed previously by Kuzmenko {\em et al.} for the bulk graphite case\cite{Kuzmenko_graphite}. 
In normal stacks, more pronounced IR features are an indicator for orthorhombically stacked sub-units.
We conclude that the optical conductivity  
or corresponding transmittance $T(\omega)=\left[1+{2\pi \over c}\sigma_R(\omega)\right]^{-2}$\cite{Kuzmenko_graphite}
spectrum can provide a convenient qualitative characterization 
of multilayer graphene stacks.  

This work was supported by the Welch Foundation, by the SWAN NRI program, and by the NSF under grant DMR-0606489.
The authors gratefully acknowledge valuable comments from M. D. Stiles, J. McClelland, J. A. Stroscio, and A. R. Hight Walker.

\end{document}